\documentstyle[epsfig,prl,aps]{revtex}
\begin{document}
\draft
\twocolumn[\hsize\textwidth\columnwidth\hsize\csname
@twocolumnfalse\endcsname

\title{Gauge-noninvariance of quantum cosmology and vacuum dark energy}
\author{Irina Dymnikova}
\address{Department of Mathematics and Computer Science,
University of Warmia and Mazury,\\
\.Zo{\l}nierska 14, 10-561 Olsztyn, Poland; e-mail: irina@matman.uwm.edu.pl}
\author{Michael Fil'chenkov}
\address{Alexander Friedmann Laboratory for Theoretical Physics,
St. Petersburg, Russia\\
Institute of Gravitation and Cosmology, Peoples' Friendship
University of Russia,\\
 6 Miklukho-Maklaya Street, Moscow 117198,
Russia; e-mail: fmichael@mail.ru}

\maketitle

\begin{abstract}

We address the question how to adapt cosmological constant
$\Lambda$ for description of a vacuum dark energy density jumping
from the big initial value to the small today value suggested by
observations. We find such a possibility in the
gauge-noninvariance of quantum cosmology which leads to a
connection between a choice of the gauge and quantum spectrum for
a certain physical quantity which can be specified in the
framework of the minisuperspace model. We introduce a particular
gauge in which the cosmological constant $\Lambda$ is quantized
and show that making a measurement of $\Lambda$ today one can find
its small value with the biggest probability, while at the
beginning of the evolution, the biggest probability corresponds to
its biggest value. Transitions between quantum levels of $\Lambda$
in the course of the Universe evolution, could be related to
several scales for symmetry breaking.

\end{abstract}


\vspace{0.17cm} ]

{\bf Introduction -} Astronomical data provide the convincing
evidence that our Universe is dominated in 70\% of its density by
a dark energy responsible for its accelerated expansion due to
negative pressure $p=w\varepsilon; ~w<-1/3$ \cite{observations}.
Observations constrain the parameter $w$ to $w<-0.7$ with the best
fit $w=-1$ \cite{ellis,bestfit} corresponding to the cosmological
constant $\Lambda$ related to a vacuum energy density by
$\Lambda=8\pi G\varepsilon_{vac}$. But $\Lambda$ is plagued by the
problem of cosmological constant: The inflationary paradigm
requires  it to be large at the earliest evolutionary stages,
observations demand that it should be many orders of magnitude
less today, while the Einstein equations require it to be
constant, and the quantum field theory estimates an upper cutoff
on this constant by the Planck scale, the resulting zero-point
contribution to gravity is incompatible with all observational
data \cite{weinberg}.

In this Letter we show the existence of the  connection between
$\Lambda$ dynamics and the gauge-noninvariance of quantum
cosmology which could support cosmological constant $\Lambda$ as a
promising  dark energy candidate.

In quantum cosmology the whole universe is treated
quantum-mechanically  and described by a wave function $\Psi$. The
wave function of a universe
$\Psi[h_{ik}(\vec{x}),\phi_m(\vec{x})]$ is defined on the
superspace of all 3-dimensional geometries $h_{ik}(\vec{x})$ and
matter field configurations $\phi_m(\vec{x})$, and  satisfies the
Wheeler-DeWitt equation \cite{dewitt,wheeler}
$$
\hat H\Psi=0.
                                                   \eqno(1)
$$

In the minisuperspace models for quantum gravity the four-metric
is typically written in the Arnowitt-Deser-Misner formalism as
$$
ds^2=N^2(t)dt^2-h_{ij}dx^i dx^j,
                                                                    \eqno(2)
$$
where  $N(t)$ is an arbitrary lapse function.

In quantum geometrodynamics there exists the problem of  time
gauge \cite{halliwell}. In quantum cosmology the essence of this
problem is clearly seen in the  minisuperspace examples  below.

On the one hand, in the Vilenkin ansatz \cite{vilenkin} the four
line element is written as
$$
ds^2=N^2(t)dt^2-a^2(t)d\Omega_3^2
                                                                     \eqno(3)
$$
where $a(t)$ is the scale factor and $d\Omega_3^2$ is the metric
on a unit 3-sphere. For a homogeneous isotropic and closed model
with the cosmological term the action is
$$
S=\frac{1}{16\pi G}\int{d^4x\sqrt{-g}[{R} -2\Lambda]},
                                                                     \eqno(4)
$$
 and the Lagrangian reads \cite{vilenkin}
 $$
 {\cal{L}}=\frac{1}{2}N\biggl(a\biggl[1-\frac{\dot{a}^2}{N^2}\biggr]-\Lambda
 a^3\biggr).
                                                                             \eqno(5)
 $$

The canonical form of the Lagrangian is $ {\cal L}=p_a\dot a -
{\cal H} = p_a\dot a-N(t)H$ where $N(t)$ plays the role of the
Lagrange multiplier. The Lagrange equation $
\partial{\cal{L}}/{\partial N}=0$
 gives the  constraint
$$
{H}=-\frac{1}{2}\biggl[\frac{p_a^2}{a}+a-\Lambda a^3\biggr]=0.
                                                                            \eqno(6)
$$
The Hamiltonian constraint for $H$, related to the Hamiltonian of
the dynamical system ${\cal H}=N(t)H$,  does not involve the
time-dependence given by $N(t)$. As a result, the standard
quantization procedure ($p_a \rightarrow -id/da$) applied to the
constraint (6), gives the Wheeler-DeWittt equation $\hat{
H}\Psi=0$ in the form independent of the lapse function $N(t)$
\cite {vilenkin} (which is out of configuration space).

On the other hand, it was emphasized by Halliwell that the
Hamiltonian constraint expresses an invariance of the theory under
time reparametrization, but the formalism is not invariant under
field redefinition involving the lapse function
$N(t)$\cite{halliwell}.

 Halliwell studied a class of reparametrization-invariant theories
 described by an action of the form
 $$
S=\int{dt[p_a {\dot q}^a-N{ H}(p_a, q^a)]}
                                                                           \eqno(7)
$$
with a Lagrange multiplier $N$ which enforces the constraint ${
H}=0$. Two  minisuperspace models, first with the  line element
$$
ds^2=-\frac{N^2(t)}{q(t)}dt^2+q(t)d\Omega_3^2
                                                                           \eqno(8)
$$
and the action (the case $k=0$)
$$
S=\int{dt\biggl[p\dot{q}-\frac{N}{2}\biggl(-4p^2-1\biggr)\biggr]},
                                                                         \eqno(9)
 $$
and second, with the lapse function $ \tilde{N}=q^{-1}N$,  the
line element
$$
ds^2=q(t)[{\tilde N}^2(t)dt^2-d\Omega_3^2]
                                                                        \eqno(10)
 $$
and the action
$$
S=\int{dt\biggl[p\dot{q}-\frac{\tilde{N}}{2}\biggl(-4qp^2-q\biggr)\biggr]},
                                                                         \eqno(11)
$$
give two different Wheeler-DeWitt equations and describe different
quantum theories, so that the procedure is not invariant under
field redifinitions involving the lapse function \cite{halliwell}
(see also \cite{shestakova}).

Indeed, as early as in 1967  DeWitt found, in the synchronous
gauge ($N=1$), the energy spectrum of a dust-filled universe in
the closed FRW model with a curvature-generated potential and
without $\Lambda$, which corresponds to quantization in the well
with infinite walls \cite{dewitt}. Adding cosmological term
$\Lambda g_{\mu\nu}$ results in transformation of an infinite well
into a finite barrier \cite{KM}. Existence of a finite barrier
allows one to consider emergence  of a universe in the quantum
tunnelling event \cite{vilenkin82}. In the synchronous gauge a
universe would emerge with a certain quantized value of the
rest-mass energy. In the conformal gauge ($N=a$) the quantized
quantity taking the role of energy in the WDW equation is related
to the contribution of the radiation ($p=\varepsilon/3$) to the
total energy  density \cite{F5}, which makes possible a quantum
birth of a hot universe in the tunnelling event \cite{us2002}.

Here we study consequences of gauge-noninvariance of the theory
under field redefinitions involving the lapse function $N$, to
$\Lambda$ dynamics.

In quantum cosmology models a universe could pass through stages
on which the cosmological constant $\Lambda$ can take different
values in a wide range \cite{weinberg}. In particular, a universe
could start from a quantum state in which $\Lambda$ had not  a
certain value; any measurements of the universe properties would
give then a distribution of possible values of $\Lambda$ with a
priori probabilities defined by an initial state \cite{hawking}.

Following these ideas, we admit that at each stage of its
evolution a universe could be found, with a certain probability,
in a state with a certain quantum value of $\Lambda$.

We show that the gauge-noninvariance of quantum cosmology leads to
a connection between a choice of the gauge function $N$ and
quantum spectrum for a certain physical quantity, which we specify
in the framework of the minisuperspace model. We introduce a
particular gauge in which the cosmological constant $\Lambda$ is
quantized. Then the wave function of the universe is
$\Psi=\displaystyle\sum_n{c_n\psi_n}$ which allows one to estimate
the probability to find the universe in a state with a certain
quantized value $\lambda_n$.

\vskip0.1in

 {\bf Approach -}
We express the line element in the form
$$
ds^2=N^2(a)d\eta^2-a^2(\eta)d\Omega^2_3
                                                                     \eqno(12)
$$
which explicitly takes into account the field redefinitions
involving the lapse function. The time-gauge function $N(a)$
enters into the configuration space (as a function of $a$), and
the dynamical system becomes clearly non-invariant under the gauge
transformations $N(a)\rightarrow \tilde{N}(\tilde{a})$.

In fact (12) represents  a generalization of the conformal gauge
$N(a)=a$ ($dt=ad\eta \rightarrow dt=N(a)d\eta$).

We start with the action
$$
S=\frac{1}{16\pi G}\int{R \sqrt{-g}d^4x}.
                                                                     \eqno(13)
$$

 Using
the freedom of adding to the Lagrangian an arbitrary full
derivative $df(\eta)/d\eta$, we express it as
$$
{\cal{L}}=\frac{1}{2}\frac{a\dot{a}^2}{N(a)}-\frac{N(a)}{2}ak
+\frac{4\pi G}{3}N(a)\varepsilon(a)a^3.
                                                                     \eqno(14)
$$
Here dot denotes differentiation with respect to $\eta$, and $k$
is the curvature parameter ($k=-1, 0, +1$ for an open, flat and
closed model respectively).

The momentum is given by
$$
p_a=\frac{a\dot{a}}{N(a)}.
                                                                     \eqno(15)
$$
The Hamiltonian reads
$$
{\cal{H}}=\frac{N(a)}{2}\frac{p_a^2}{a}+\frac{N(a)}{2}ak
-\frac{4\pi G}{3}N(a)\varepsilon(a)a^3.
                                                                    \eqno(16)
$$
The canonical form of the Lagrangian is
$$
{\cal L}=p_a\dot a-{\cal H}=p_a\dot a -N(a)\tilde{\cal H}
                                                                   \eqno(17)
$$
where
$$
\tilde{\cal H}=\frac{1}{2}\frac{p_a^2}{a}+\frac{1}{2}ak
-\frac{4\pi G}{3}\varepsilon(a)a^3.
                                                                  \eqno(18)
$$
The gauge function $N(a)$ plays the role of the Lagrange
multiplier, which gives the constraint equation
$$
{\cal H}=0.
                                                                \eqno(19)
$$

 To present the Hamiltonian (16) in the canonical form, we make the
canonic transformation $a, p_a\rightarrow q, p_q$ such that
$$
 p_q=\sqrt{\frac{N(a)}{f(a)a}}; ~ ~~ q=\int{\sqrt{\frac{af(a)}{N}}da}
                                                                  \eqno(20)
$$
where $f(a)$ is a smooth function. It is easy to check that the
second Hamilton equation $\dot{p_q}=-\partial{\cal{H}}/\partial q$
is satisfied identically for any function $f(a)$, so we can put
$f=1$. The resulting Hamiltonian reads
$$
{\cal{H}}=\frac{p^2_q}{2} +\frac{k}{2}N(q)a(q)-\frac{4\pi
G}{3}\varepsilon(a(q))a^3(q)N(q).
                                                                    \eqno(21)
$$
The standard procedure of quantization ($p_q\rightarrow
-i\frac{d}{dq}$) gives the Wheeler-DeWitt equation \footnote{To
allow for an operator ordering the kinetic term should be
represented in the form
$\frac{1}{q^p}\frac{d}{dq}q^p\frac{d\Psi}{dq}$. Here we adopt for
simplicity $p=0$ following \cite{vilenkin,vilenkin82}.}
$$
\left(-\frac{d^2}{dq^2} + V(q)\right)\Psi(q)=0
                                                                 \eqno(22)
$$
where
$$
V(q)=\frac{1}{l_{Pl}^4}\biggl(N(a(q))ka(q)-\frac{8\pi
G}{3}\varepsilon(a(q))a^3(q)N(q)\biggr)
                                                                     \eqno(23)
$$
The energy density $\varepsilon(a)$ can be written in the form
\cite{F5}
$$
\varepsilon=\sum_{n} B_n a^{-n}.
                                                                       \eqno(24)
$$
The coefficients $B_n$ refer to contributions of different
non-interacting components of the matter content. The parameter
$n$ is connected by $ n=3(1+\alpha)$ with the parameter $\alpha$
in the equation of state $p=\alpha\varepsilon$.

A choice of the gauge $a^{3-n}N(a)=l_{Pl}^{4-n}$ separates a
scale-factor-free term in the potential (23), as a result, the
Wheeler-DeWitt equation (22) reduces to
$$
\frac{{\hbar}^2}{2m_{Pl}}\frac{d^2\Psi}{dq^2}
-\frac{E_{Pl}}{2}\biggl(U(a(q))-Q\biggr)\psi=0
                                                                       \eqno(25)
$$
 with the eigenvalue $Q$ given by
$$
Q=\frac{8\pi}{3}{B_n}.
                                                                       \eqno(26)
$$
Dependence $q(a)$ is given by
$$
q=\frac{1}{(3-n/2)}a^{3-n/2}
                                                                             \eqno(27)
$$
Equation (25) describes a quantum system with the quantized
quantity $Q$ related to the contribution of the matter component
specified by $n$, in the potential $U(q)$ created by other
components of the matter content.

An appropriate choice of the boundary conditions is the De Witt
boundary condition $\Psi=0$ at $a(q)=0$ \cite{dewitt}, and the
Vilenkin boundary condition \cite{vilenkin2} - an outgoing wave
function out of a barrier.

\vskip0.1in

{\bf Quantization of $\Lambda$ -} In the gauge $N(a)a^3=1$
\footnote{At the classical level this gauge corresponds to such a
theory in which $\Lambda$ appears as a constant of integration in
the Einstein equations without cosmological term \cite{weinberg}.}
our quantum system is described  by the Schr\"odinger equation
(written in the Planckian units)
$$
\frac{d^2\Psi}{dq^2}-(U(q)-Q_{\Lambda})\Psi=0
                                                                       \eqno(28)
$$
where
$$
Q_{\Lambda}=\frac{8\pi}{3}B_0=
\frac{8\pi}{3}\varepsilon_{vac}=\frac{\Lambda}{3}.
                                                                      \eqno(29)
$$
Equation (28) describes a quantum system with the quantized
quantity $\Lambda$ (vacuum density) in the potential $U(q)$
generated by other matter components which include radiation,
dust, and some admixture of strings or quintessence with the
equation of state $p=-\varepsilon/3$. The latter does not affect
acceleration but contribute to the curvature term
\cite{kardashev,F5,us2001} which facilitates quantum birth of an
open and flat universe \cite{F5,us2002,us2001}. In this case
$$
U(a)=\frac{(k-B_s)}{a^2}-\frac{B_d}{a^3}-\frac{B_{\gamma}}{a^4};
 ~~ ~a^3=3q
                                                                  \eqno(30)
$$
with $ B_s=\frac{8\pi}{3}B_2;~~B_d=\frac{8\pi}{3}B_3;~~
B_{\gamma}=\frac{8\pi}{3}B_4$. In the presence of the curvature
term $(k-B_s)/a^2$ the potential $U(a(q))$ represents a barrier
shown in Fig.1.
\vskip0.1in
\begin{figure}
\vspace{-8.0mm}
\begin{center}
\epsfig{file=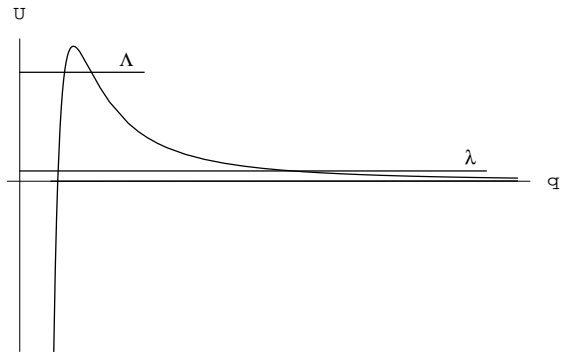,width=8.0cm,height=5.0cm}
\end{center}
\caption {Typical behavior of a potential $U(q)$. Two relevant
(remote past and present) quantum levels of $\Lambda$ are shown.}
\label{fig.1}
\end{figure}
 The potential $U(q)$ has one zero, one maximum,  goes to
minus infinity as $q\rightarrow 0$, and vanishes at plus infinity,
so that negative eigenvalues of $\Lambda$ are confined.

The wave function is the superposition of states
$$
\Psi=\displaystyle\sum_{n}a_n\psi_n
                                                                       \eqno(31)
$$
where the eigenfunctions satisfy
$$
\frac{d^2\psi_n}{dq^2}-(U(q)-\lambda_n)\psi_n=0
                                                                      \eqno(32)
$$
with the eigenvalues
$$
\lambda_n=\frac{8\pi}{3}\varepsilon_{vac}.
                                                                     \eqno(33)
$$
The quasiclassical solution to (32) in the region outside of a
barrier is given by
$$
\psi_n=\frac{1}{\sqrt{\lambda_n-U(q)}}e^{\pm
i\int{\sqrt{\lambda_n-U(q)}}dq}
                                                                  \eqno(34)
$$
Qualitative behavior of the probability near the maximum of the
potential $U_m=U(a_m)$
$$
|\psi_n|^2\simeq{\frac{1}{\lambda_n-U_m+|U^{\prime\prime}(a_m)|(a-a_m)^2/2}}
                                                                \eqno(35)
$$
corresponds to domination of a big eigenvalue $\lambda_n$.

Behavior far from the maximum of the potential is
$$
|\psi_n|^2\simeq{\frac{a^4}{\lambda_n a^4-(k-B_s)a^2 + B_d a
+B_{\gamma}}}
                                                                \eqno(36)
$$
and the probability is maximal near an intersection point
$\lambda_n=U(q)$ where some $\lambda_n$ starts to dominate at some
vacuum-dominated stage.

The picture of $\Lambda$ dynamics looks as follows. Making a
measurement of $\lambda_n$ today, we find its small value
$\lambda$ with the big probability. At the beginning of evolution,
the biggest probability corresponds to the biggest value
$\Lambda$. Evolution involving transitions between quantum levels
$\lambda_n$ can be related to several scales for symmetry
breaking.

\vskip0.1in

{\bf Discussion-}
 The aim of this Letter was to note the existence in principle of the
possibility to adapt a cosmological constant for description of a
dark energy, rather than to give precise estimates in the detailed
model.

As we have seen in the framework of the minisuperspace model, the
gauge non-invariance of quantum cosmology provides the possibility
for quantization of different physical quantities, in particular,
of cosmological constant $\Lambda$ associated with the vacuum
density $\rho_{vac}$.

Such a possibility implies that a potential in the Wheeler-DeWitt
equation represents a finite barrier. In the case of quantized
$\Lambda$  this suggests the presence in the matter  content a
component with the equation of state $p=-\varepsilon/3$ which
could be some kind of Q-matter, e.g., strings with the negative
deficit angle \cite{us2002}.

Indeed, the parameter $k-B_s$ at the start of the present
vacuum-dominated epoch with $\lambda_n=\lambda$ can be estimated
at the maximum of the probability (36) as
$$
k-B_s \simeq{\lambda a^2 + \frac{B_d}{a}}
                                                             \eqno(37)
$$
where $\lambda$ corresponds to the vacuum density today
$\rho_{vac}\simeq{0.7\rho_{tot}}$ \cite{observations}. According
to observations, cosmological vacuum responsible for the present
inflation starts to dominate at the age $2\times 10^9$ years
\cite{domin}. The rough estimate of $\rho_s$ from (37) suggests,
for the case of the vacuum-dominated spatially flat universe
($k=0$), an admixture of a matter component with the equation of
state $p=-\varepsilon/3$, at the level
$\rho_s\sim{10^{-2}\rho_{tot}}$.

 \vskip0.1in

{\bf Acknowledgement}

This work was supported by the Polish Ministry of Science and
Information Society Technologies through the grant 1P03D.023.27.

\end{document}